\documentclass[runningheads,a4paper]{llncs}
\usepackage{amssymb}
\usepackage{graphicx}
\usepackage{url}
\newcommand{\keywords}[1]{\par\addvspace\baselineskip
\noindent\keywordname\enspace\ignorespaces#1}
\begin{document}

\mainmatter

\graphicspath{{images/}}

\title{Francy - An Interactive Discrete Mathematics Framework for GAP}  
\titlerunning{Francy} 
\author{Manuel Machado Martins\inst{1} \and Markus Pfeiffer\inst{2}}
\authorrunning{Martins-Pfeiffer}
\institute{
Universidade Aberta, Portugal\\
\email{manuelmachadomartins@gmail.com},\\ 
\texttt{https://github.com/mcmartins}
\and
University of St Andrews, Scotland\\
\email{markus.pfeiffer@st-andrews.ac.uk},\\ 
\texttt{https://markusp.morphism.de}
}
\maketitle

\begin{abstract}
Data visualization and interaction with large data sets is known to be essential and critical in many businesses today, and the same applies to research and teaching, in this case, when exploring large and complex mathematical objects.
GAP is a computer algebra system for computational discrete algebra with an emphasis on computational group theory. The existing XGAP package for GAP works exclusively on the X Window System. It lacks abstraction between its mathematical and graphical cores, making it difficult to extend, maintain, or port. In this paper, we present Francy, a graphical semantics package for GAP. Francy is responsible for creating a representational structure that can be rendered using many GUI frameworks independent from any particular programming language or operating system.
Building on this, we use state of the art web technologies that take advantage of an improved REPL environment, which is currently under development for GAP. The integration of this project with Jupyter provides a rich graphical environment full of features enhancing the usability and accessibility of GAP.
\end{abstract}

\section{Introduction}

By providing a mechanism for quickly demonstrating a topic, or result, visual learning has been proven effective and advantageous. It helps with engagement and allows students to look at problems in a different way \cite{VisualLeap:2017:Online}. 

In mathematics, especially in group theory, having a graphical representation of certain structures is invaluable when formulating conjectures and counterexamples, and when analyzing data. GAP is a computer algebra system (CAS) focused on computational group theory and it helps to explore algebraic structures and solve a variety of problems \cite{GAP:2018:Online}. The primary existing package for GAP, which provides facilities displaying graphics and visualization of mathematical data structures, is XGAP. This package is integrated with the Unix X-Window System, which provides a basic framework for a Graphical User Interface (GUI), and includes a wide range of mathematical functionality focused on the lattice of subgroup of a group \cite{XGAPDoc:2018:Online}. Further such packages include Interactive Todd Coxeter (ITC) which was developed using XGAP and provides an interactive environment for exploring coset enumerations \cite{ITCDoc:2018:Online}. GAP.APP is another project based on XGAP and it provides a native Macintosh interface for GAP \cite{GAP.APP:2018:Online}. All of these projects enable GAP to be used as a tool to visualize objects with computer graphics.

Technology evolves quickly and today multiple web platforms allow users to experience, learn, and share in a simple and fast paced environment. Jupyter is one of these projects and, as mentioned on the official website \cite{Jupyter:2018:Online}, aims ``\emph{to develop open-source software, open-standards, and services for interactive computing across dozens of programming languages}", leveraging learning processes and the way people share their work. The purpose of the OpenDreamKit project is to provide a framework for the advancement of mathematics in Europe, as part of the Horizon2020 European Research Infrastructure, and Jupyter is a core component of OpenDreamKit \cite{OpenDreamKit:2018:Online}. Jupyter allows a centralized system for the dissemination of content and uses an intuitive interface where users interact with notebooks containing live code, equations, visualizations and narrative text.

\textsf{Francy} \cite{Francy:2018:Online} arose from the necessity of having a lightweight framework for building interactive graphics, generated from GAP, running primarily on the web, primarily in a Jupyter Notebook. An initial attempt to re-use XGAP and port it was made, but the lack of a standardized data exchange format between GAP and the graphics renderer, and the simplistic initial requirements of the project were the basis for the creation of a new GAP package.

\section{Functionality}

\textsf{Francy} provides an interface to draw graphics using objects. This interface is based on simple concepts of drawing and graph theory, allowing the creation of directed and undirected graphs, trees, line charts, bar charts and scatter charts. These graphical objects are drawn inside a canvas that includes a space for menus and to display informative messages. Within the canvas it is possible to interact with the graphical objects by clicking, selecting, dragging and zooming.

In terms of interaction with the kernel, we use callbacks which allow the execution of functions in GAP from the graphical objects. A callback holds the function signature and any arguments that it requires. If a callback requires user input, a modal window will be shown before the execution of the function.

\section{Applications}

\textsf{Francy} does not provide any mathematical functionality as it is intended to be used by other mathematical software packages. Existing GAP packages can be easily ported to use it. \textsf{Francy} has potentially many applications and can be used to provide a graphical representation of data structures, allowing one to navigate through and explore properties or relations of these structures. In this way, \textsf{Francy} can be used to enrich a learning environment where GAP provides a library of thousands of functions implementing algebraic algorithms as well as large data libraries of algebraic objects.

In the following code we show a simple usage of \textsf{Francy} to display interactively the directed graph of all subgroups of the Symmetric Group $S_3$, using the GAP package Digraphs \cite{Digraphs:2018:Online}:

\begin{verbatim}
LoadPackage("digraphs"); LoadPackage("francy");

G := SymmetricGroup(3); as := AllSubgroups(G); nodes := [];
d := Digraph(as, {H, K} -> IsSubgroup(H, K));

vertices := DigraphVertices(d); edges := DigraphEdges(d);

canvas := Canvas(Concatenation("Subgroups Digraph of ", 
  String(G)));
graph := Graph(GraphType.DIRECTED); 
Add(canvas, graph);

customMessage := FrancyMessage(FrancyMessageType.INFO, 
  "Simple Groups", "A group is simple if it is nontrivial 
  and has no nontrivial normal subgroups.");

IsGroupSimple := function(i)
  Add(canvas, simpleGroupMessage);
  if IsSimpleGroup(as[i]) then
    Add(canvas, FrancyMessage("Simple", 
      Concatenation("The vertex ", String(i),
        ", representing the subgroup ", String(as[i]), 
        ", is simple.")));
  else
    Add(canvas, FrancyMessage("Not Simple",
      Concatenation("The vertex ", String(i),
        ", representing the subgroup ", String(as[i]), 
        ", is not simple.")));
  fi;
  return Draw(canvas);
end;

for i in vertices do
  nodes[i] := Shape(ShapeType.CIRCLE, String(i));
  Add(nodes[i], Menu("Is this subgroup simple?", 
      Callback(IsGroupSimple, [i])));
  Add(graph, nodes[i]);
od;

for i in edges do 
  Add(graph, Link(nodes[i[1]], nodes[i[2]]));  
od;

Draw(canvas);
\end{verbatim}

\begin{figure}
  \begin{center}
  \includegraphics[scale=0.5]{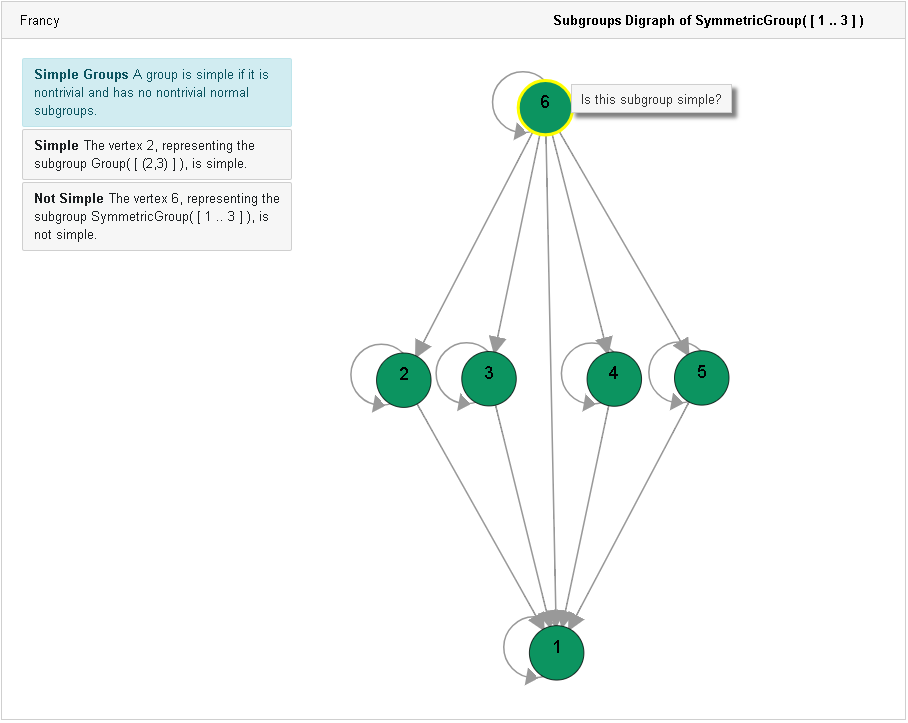}
  \caption{The graphics produced by the code listing above.}
  \end{center}
\end{figure}

\section{Technical contribution}

In terms of software design, \textsf{Francy} follows some principles such as Separation of Concerns and Modularity. These principles are perfectly articulated in the Computer Science Handbook \cite{ComputerScience:2004:Online} ``\emph{Any domain or application can be divided and decomposed into major building blocks and components (separation of concerns). This decomposition allows the application requirements to be further defined and refined, while partitioning these requirements into a set of interacting components (modularity). Changes to the application are (it is hoped) localized. In addition, team-oriented design and development can proceed with different team members concentrating on particular components}". 

\textsf{Francy} consists of two main components, a GAP package that is responsible for the semantic representation of graphics, and a second component, a GUI library that is responsible for generating the actual interactive graphical representation.

The GAP package creates a semantic representation of graphics, providing a thin layer between GAP objects and graphical objects to be rendered. This is done using JSON, a lightweight, text-based, language-independent data interchange format \cite{JSON:2017:Online}. The semantic model follows a JSON Schema \cite{FrancySchema:2018:Online,JSONSCHEMA:2017:Online}, and is identified with the \textit{application/vnd.francy+json} MIME type \cite{MIMEType:2018:Online}. This creates an abstraction and allows the development of new GUI libraries, using different data rendering dependencies or even different programming languages, independently of the GAP package. This package is somehow based in XGAP throughout its application programming interface (API), but avoiding any non-GAP code. This has been the main concern, in order to allow a smooth integration with other GAP packages. In fact, \textsf{Francy} has only one dependency, the JSON package \cite{JSONGAP:2017:Online}, that is distributed with GAP by default, and it is needed to communicate with Jupyter. Access to the GAP language shell (Read–Eval–Print Loop or REPL) is abstracted and managed by a kernel \cite{JupyterKernelGateway:2018:Online,GAPJupyterKernel:2018:Online}.

At the moment, \textsf{Francy} has a JavaScript GUI library, based on d3.js \cite{D3JS:2018:Online}, for rendering the semantic representation produced by the GAP package. This library is distributed both as a browser module and as a Jupyter extension. The browser module can be used for displaying graphics outside a Jupyter environment or to build applications that can be integrated with GAP, for instance, using WebSockets \cite{WebSockets:2018:Online} and a web-based terminal emulator such as tty.js \cite{TTY.JS:2018:Online}. The Jupyter extension can be used in Jupyter Notebooks or Jupyter Lab, using the Jupyter GAP Kernel \cite{GAPJupyterKernel:2018:Online} and the MIME type \textit{application/vnd.francy+json} to render the document. 

\section{Future work}

Many other interactive features can be implemented providing a richer learning environment. Features such as rendering multiple topological graphs on the same canvas would allow, for instance, easier comparison of data structures. Other ways for users to input data would provide a more intuitive user experience. 

Packages such as the Francy-Monoids \cite{FrancyMonoids:2018:Online}, Subgroup Lattice \cite{SubgroupLattice:2018:Online} and the Interactive Todd-Coxeter \cite{InteractiveToddCoxeter:2018:Online} still need to be polished and finished.

At the moment, the semantic model based is not being validated against the JSON Schema \cite{FrancySchema:2018:Online}. This can be addressed in the future by extending the actual JSON package and implement the JSON Schema specification for validation of documents.

It would also be beneficial moving some of the processing JavaScript code into Web Workers \cite{WebWorkers:2018:Online}, such that rendering of huge structures does not block the web page.

In some cases, having a local installation of \textsf{Francy} could be a requirement, and porting it to a desktop application is also possible as there are many tools to help on this process, for instance with ElectronJS \cite{ElectronJS:2018:Online}.

\section*{Acknowledgements}

We are grateful to James D. Mitchell, Pedro A. Garc\'ia-S\'anchez, Jo\~ao Ara\'ujo and Francesca Fusco for their suggestions that led to a much improved version of the paper.

We are also very grateful to the anonymous referees for their careful reviews and helpful suggestions.

The first author is grateful to CoDiMa (CCP in the area of Computational Discrete Mathematics - EPSRC EP/M022641/1, 01/03/2015-29/02/2020) for supporting the attendance at the event Computational Mathematics with Jupyter 2017 in Edinburgh, in which some of this research was done. The second author has received funding from the European Union project Open Digital Research Environment Toolkit for the Advancement of Mathematics (EC Horizon 2020 project 676541, 01/09/2015-31/08/2019).

\bibliography{biblio}

\begin{thebibliography}{10}

\bibitem{VisualLeap:2017:Online}
J.~Berg.
\newblock {\em Visual Leap: A Step-by-Step Guide to Visual Learning for
  Teachers and Students}.
\newblock Bibliomotion, Incorporated, 2015.

\bibitem{GAP:2018:Online}
The~GAP Group.
\newblock Gap - groups, algorithms, and programming, version 4.9.1.
\newblock \url{https://www.gap-system.org/}, 2018.

\bibitem{XGAPDoc:2018:Online}
Max~Neunh\"offer Frank~Celler.
\newblock Xgap documentation, what is xgap?
\newblock
  \url{https://www.gap-system.org/Manuals/pkg/xgap-4.26/htm/CHAP002.htm}, 2018.

\bibitem{ITCDoc:2018:Online}
Joachim~Neub\"user Volkmar~Felsch, Ludger~Hippe.
\newblock {ITC} documentation", what is {ITC}?
\newblock \url{https://www.gap-system.org/Manuals/pkg/itc/htm/CHAP001.htm},
  2018.

\bibitem{GAP.APP:2018:Online}
Russ Woodroofe.
\newblock Introducing gap.app.
\newblock \url{https://cocoagap.sourceforge.io/}, 2018.

\bibitem{Jupyter:2018:Online}
Jupyter Community.
\newblock Project jupyter.
\newblock \url{http://jupyter.org/}, 2018.

\bibitem{OpenDreamKit:2018:Online}
OpenDreamKit Community.
\newblock Project opendreamkit.
\newblock \url{http://opendreamkit.org/}, 2018.

\bibitem{Francy:2018:Online}
Manuel~Machado Martins.
\newblock Francy github page.
\newblock \url{https://github.com/mcmartins/francy/}, 2018.

\bibitem{Digraphs:2018:Online}
Jan~De Beule, Julius Jonu{\v s}as, James~D. Mitchell, Michael Torpey, and
  Wilf~A. Wilson.
\newblock Digraphs - gap package, version 0.12.1, Apr 2018.
\newblock doi: \url{10.5281/zenodo.596465}.

\bibitem{ComputerScience:2004:Online}
A.B. Tucker.
\newblock {\em Computer Science Handbook, Second Edition}.
\newblock CRC Press, 2004.

\bibitem{JSON:2017:Online}
Douglas Crockford.
\newblock The javascript object notation (json) data interchange format.
\newblock \url{https://tools.ietf.org/html/rfc8259}, 2018.

\bibitem{FrancySchema:2018:Online}
Manuel~Machado Martins.
\newblock Francy schema github page.
\newblock
  \url{https://github.com/mcmartins/francy/blob/master/gap/schema/francy.json},
  2018.

\bibitem{JSONSCHEMA:2017:Online}
Json~Schema Community.
\newblock Json schema.
\newblock \url{http://json-schema.org/}, 2018.

\bibitem{MIMEType:2018:Online}
Mozilla and individual contributors.
\newblock Mime types.
\newblock
  \url{https://developer.mozilla.org/en-US/docs/Web/HTTP/Basics_of_HTTP/MIME_types},
  2018.

\bibitem{JSONGAP:2017:Online}
Christopher Jefferson.
\newblock json - reading and writing json.
\newblock
  \url{https://www.gap-system.org/Manuals/pkg/json-1.2.0/doc/chap0.html}, 2018.

\bibitem{JupyterKernelGateway:2018:Online}
Jupyter Community.
\newblock Jupyter documentation, jupyter kernel gateway.
\newblock \url{http://jupyter-kernel-gateway.readthedocs.io/en/latest/}, 2018.

\bibitem{GAPJupyterKernel:2018:Online}
Markus Pfeiffer.
\newblock Native jupyter kernel for gap github page.
\newblock \url{https://github.com/gap-packages/JupyterKernel}, 2018.

\bibitem{D3JS:2018:Online}
Mike Bostock.
\newblock Data-driven documents, d3.
\newblock \url{https://d3js.org/}, 2018.

\bibitem{WebSockets:2018:Online}
Mozilla and individual contributors.
\newblock Websockets.
\newblock
  \url{https://developer.mozilla.org/en-US/docs/Web/API/WebSockets_API}, 2018.

\bibitem{TTY.JS:2018:Online}
Christopher Jeffrey.
\newblock tty.js github page.
\newblock \url{https://github.com/chjj/tty.js}, 2018.

\bibitem{FrancyMonoids:2018:Online}
Pedro~A. Garc\'ia-S\'anchez.
\newblock Francy monoids github page.
\newblock \url{https://github.com/pedritomelenas/francy-monoids}, 2018.

\bibitem{SubgroupLattice:2018:Online}
Manuel~Machado Martins.
\newblock Subgroup lattice github page.
\newblock \url{https://github.com/mcmartins/subgroup-lattice}, 2018.

\bibitem{InteractiveToddCoxeter:2018:Online}
Manuel~Machado Martins.
\newblock Interactive todd-coxeter github page.
\newblock \url{https://github.com/mcmartins/interactive-todd-coxeter}, 2018.

\bibitem{WebWorkers:2018:Online}
Mozilla and individual contributors.
\newblock Web workers api.
\newblock
  \url{https://developer.mozilla.org/en-US/docs/Web/API/Web_Workers_API}, 2018.

\bibitem{ElectronJS:2018:Online}
Electron Community.
\newblock Build cross platform desktop apps with javascript, html, and css.
\newblock \url{https://electronjs.org/}, 2018.

\end{thebibliography}
\bibliographystyle{unsrt}
\end{document}